\documentclass[twocolumn,superscriptaddress,preprintnumbers,pra]{revtex4-2}

\usepackage{graphicx}
\usepackage{dcolumn}
\usepackage{bm}
\usepackage{braket} 
\usepackage{amsthm,amsmath,amssymb}
\usepackage{mathrsfs}
\usepackage{graphicx}
\usepackage{float}

\usepackage{hyperref}
\hypersetup{hypertex=true,
            colorlinks=true,
            linkcolor=blue,
            anchorcolor=blue,
            citecolor=blue}

\begin{document}
\title{Optimized nonadiabatic holonomic quantum computation via reverse engineering} 

\author{Yue-Heng Liu}
\affiliation{GBA Branch of Aerospace Information Research Institute, Chinese Academy of Sciences, Guangzhou 510535, China}
\affiliation{Guangdong Provincial Key Laboratory of Terahertz Quantum Electromagnetics, Guangzhou 510700, China}

\author{Qi Li}
\email{liqi@aircas.ac.cn}
\affiliation{GBA Branch of Aerospace Information Research Institute, Chinese Academy of Sciences, Guangzhou 510535, China}
\affiliation{Guangdong Provincial Key Laboratory of Terahertz Quantum Electromagnetics, Guangzhou 510700, China}

\date{\today}

\begin{abstract}
The challenge in building high-fidelity quantum gates lies in overcoming control errors and decoherence effects caused by the coupling between the quantum system and the external environment. Nonadiabatic holonomic quantum computation uses the topological protection of the cyclic evolution of the computational subspace to make holonomic gates highly robust to control errors. Therefore, our main goal is to accelerate this evolution. Here we propose a general reverse engineering approach to combine the unconventional geometric quantum computation with optimized holonomic quantum computation [Bao-Jie Liu et al. \href{https://journals.aps.org/prl/abstract/10.1103/PhysRevLett.123.100501}{Phys.Rev.Lett.123,100501 (2019)}]. Our approach allows us to select evolution paths that require less time. Consequently, the proposed scheme is highly flexible and promising for achieving robust quantum computation in the future.
\end{abstract}

\maketitle
\section{Introduction}
Quantum computation, due to the coherence and entanglement of the quantum process, has shown many advantages in solving some hard problems\cite{nielsen2010quantum}, such as factorization of large prime numbers\cite{shor1999polynomial} and large-scale searching\cite{grover1996fast}. In order to construct a quantum computer, high-fidelity gates are necessary. However, the performance of quantum gates is generally limited by system errors and decoherence induced by the environment.
The robustness of quantum evolution presents a critical challenge that must be solved. Consequently, finding solutions to the robustness issues in quantum evolution is an imperative for the field.

The geometric phase depends only on the shape of the closed evolution path and is independent of the details of the evolution process\cite{berry1984quantal,anandan1990geometry,zanardi1999holonomic,sjoqvist2012non}. Therefore, the geometric phase has intrinsic tolerance to certain local control noises. Hence, using geometric phases to construct quantum gates naturally becomes a promising fault-tolerant method. According to the dimension of the geometric phase, geometric strategies can be divided into Abelian geometric quantum computation\cite{zhao2017rydberg,zhu2002implementation} and non-Abelian holonomic quantum computation (HQC)\cite{leroux2018non}. In early HQC, the adiabatic condition was necessary to prevent information leakage from the computational space to the non-computational space. However, the adiabatic condition means that a long evolution time is required, which is not conducive to the robustness of the quantum gate. Subsequently, non-adiabatic holonomic quantum computation (NHQC) was proposed\cite{sjoqvist2012non}, in which the adiabatic condition was relaxed to the requirement that the computational subspace needs to undergo a cyclic evolution. Since then, NHQC has been widely used in theoretical and experimental research in quantum computation\cite{shen2023accelerated,shen2021ultrafast,xu2024dynamically,xu2024chiral,feng2013experimental,xu2012nonadiabatic,xu2018path,herterich2016single,su2024nonadiabatic,xing2021realization,xu2018single,li2017experimental,zhu2019single}.\par
Nevertheless, a significant problem must be addressed in geometric quantum computation: how to deal with the non-robust dynamical phase that accumulate,  Researchers have proposed different ways to solve this problem. A widely accepted approach is to completely eliminate the dynamical phase by controlling the pulse design and path rotation\cite{ding2021path,ji2021noncyclic,li2021dynamically}. The orange-slice-shaped path (OSSP), where the evolution path is enclosed by two longitude geodesics, is a very conventional choice in this approach\cite{ji2021noncyclic,li2021dynamically,xu2015nonadiabatic,hong2018implementing}. However, this choice will lead to a longer evolution path for the quantum state, which means that under the same pulse control, the evolution time will be increased. The proposal of unconventional geometric quantum computation\cite{zhu2003unconventional,chen2020robust} can effectively mitigate this problem. The central idea of the unconventional geometric quantum computation scheme is that for certain quantum evolutions, we can also implement fault-tolerant computation by using the total phase accumulated in the evolution, provided that it depends only on global geometric features of the evolution. In other words, the ratio between the dynamical phase and the geometric phase should be independent of the system parameters\cite{zhao2016nonadiabatic,hong2024unconventional,liu2023optimized}.\par
There is also a very strict constraint in NHQC\cite{liu2019plug}, which weakens its flexibility and increases the difficulty of designing control pulses and evolution paths. The dynamical and geometric phases in NHQC are high-dimensional matrices. In order to eliminate the dynamical phase, we must make every element of the dynamical matrix equal to 0. There is no doubt that, except for some simple evolution paths such as OSSP, this condition is difficult to achieve. A method is proposed in the literature to mutually cancel the off-diagonal elements of the dynamical and geometric matrices, thereby achieving a net zero contribution from them. In this case, we only need to make the diagonal elements of the dynamical matrix strictly equal to 0 to achieve fault-tolerant computation. This method is so called NHQC+\cite{liu2019plug,liu2022optimized}.\par
Finally, the cyclic evolution of quantum space makes NHQC effectively resistant to control noise. In addition, it is also necessary to consider the decoherence effect caused by the coupling between the physical system and the external environment\cite{nielsen2010quantum}. This implies that we not only need to design the Hamiltonian to ensure the computational space evolves cyclically and eliminates the dynamic phase, but also require the evolution time to be sufficiently short. For example, OSSP is the most convenient way to satisfy the first two conditions, as it enables the quantum state to evolve along the geodesic, thereby eliminating the dynamical phase.
However, a major limitation of this approach is that, regardless of how small the rotation angle of the quantum gate is, the path length is always $2\pi$. In other words, the required time is always about $2\pi/\Omega$ when a square wave is used. No general method currently exists to systematically derive such Hamiltonians that minimize the evolution time.

\par
In this paper, the unconventional geometric scheme and NHQC+ are combined for the first time. We utilize NHQC+ to enhance the flexibility of quantum gate construction, while employing the unconventional geometric scheme to shorten the evolution time. However, this necessitates a more complex pulse design than previously required. To address this, we propose a general reverse engineering method to achieve the desired Hamiltonian. Our approach enables the derivation of the target Hamiltonian and selects evolution paths that costs less time.
\section{general model}
The holonomic quantum computation\cite{sjoqvist2012non} describes that the position of $L$-dimensional subspace $\mathcal{S}_L(t)$ spanned by $\{\psi_i(t)\}_{i=1}^L$ returns 
to the initial configuration after time $T$, where $T$ is the evolution period. $\{\psi_i(t)\}$ is the solution of the Schrödinger equation, i.e., $i\partial_t\ket{\psi_i(t)}=H(t)\ket{\psi_i(t)}$.\par
The above physical process can be expressed as $\mathcal{S}_L(0)=\mathcal{S}_L(T)$. This is equivalent to stating that the initial and final spaces are both spanned by the same set of orthogonal normalized bases.
We denote this set of auxiliary bases as $\{v(t)\}_{i=1}^{L+1}$ with $\ket{v_i(0)}=\ket{v_i(T)}=\ket{\psi_i(0)}$, which need not be the solutions of the Schrödinger equation. The evolution states $\{\psi_i(t)\}_i^{L+1}$ can be expanded
\begin{eqnarray}
    &&\ket{\psi_i(t)}= \sum_{j=1}^L C_{ji}(t)\ket{v_j(t)}\label{eq:1}\\
    &&\ket{\psi_{L+1}(t)}= e^{i\gamma(t)}\ket{v_{L+1}(t)}.\label{eq:2}
\end{eqnarray}
Substituting Eq.~(\ref{eq:1}) into the Schrödinger equation yields the form of the matrix $C$,
\begin{equation}
\partial_t C_{ji}(t) = i\sum_{k=1}^L\left[A_{jm}(t)-K_{jm}(t)\right]C_{mi}(t),\label{eq:3}
\end{equation}
where $A_{ji}(t)=i\braket{v_j(t)|\partial_t|v_i(t)}$  is the geometric part,  $K_{ji}(t)=\braket{v_j(t)|H(t)|v_i(t)}$ is the dynamical part,  and $\gamma(t)$ is a real function of $t$ with $\gamma(0)=0$. Eq.~(\ref{eq:3}) can be solved to obtain
\begin{equation}
    C(t)= \mathcal{T} exp\left[i\int_{0}^{t} A(t')-K(t')dt'\right]. 
\end{equation}
In conventional holonomic quantum computation, we require that $K_{ji}(t')=0$ to ensure pure geometric evolution. The disadvantage of this approach is that, to compensate for the absence of the dynamic phase, the geometric phase must be increased. This, in turn, corresponds to the evolution path of the quantum state becoming longer, resulting in a longer evolution time.
To solve this problem, Ref.\cite{liu2019plug} introduced the NHQC+, where the off-diagonal elements of the matrices $A$ and matrix $K$ cancel each other out, i.e.,
$A_{ji}= K_{ji}$ for $i\neq j$.  Therefore, the only requirement for NHQC+ is that the diagonal elements of the matrix $K$ satisfy $K_{ii}=0$.\par
The first goal of our work is to combine NHQC+ and unconventional geometric quantum computation to further accelerate geometric gates and significantly enhance the flexibility of pulse design. . We denote this method as UNHQC+. In the unconventional geometric scheme, we no longer require the dynamic phase to be zero. The holonomic matrix $A$ and dynamical matrix $K$ in UNHQC+ are necessary to satisfy that
\begin{eqnarray}
    K_{ji}&&=A_{ji} , i\neq j\label{eq:5}\\
    K_{ji}&&= -\eta A_{ji}, i=j\label{eq:6},
\end{eqnarray}
where $\eta$ is a real proportional constant independent on parameters of the qubit system. Eq.~(\ref{eq:6}) ensure that the diagonal elements of the matrix $K$
have the geometric feature.
UNHQC+ imposes additional constraints, meaning that more complex quantum control is required. The second innovation of this work is the proposal of a general method to solve this problem. When the UNHQC+ constraints Eq.~(\ref{eq:5}) and Eq.~(\ref{eq:6}) were satisfied. Eq.~(\ref{eq:3}) can be rewritten as 
\begin{eqnarray}
    \partial_t C_{ji}(t) =&& i\sum_{m=1}^L\left[A_{jm}(t)-K_{jm}(t)\right]C_{mi}(t)\delta_{jm}\nonumber\\
                        =&& i\left[(1+\eta)A_{jj}(t)\right]C_{ji}(t).\label{eq:7}
\end{eqnarray}
Since the states $\{\psi_i(t)\}_{i=1}^{L+1}$ are the solutions to the Schrödinger equation, the Hamiltonian can be written as:
\begin{equation}
    H(t)=i\sum_{i=1}^{L+1}\ket{\dot\psi_i(t)}\bra{\psi_i(t)}.\label{eq:8}
\end{equation}
Substituting Eq.~(\ref{eq:1}) and Eq.~(\ref{eq:2}) into Eq.~(\ref{eq:8}), the general form of engineering for UNHQC+ can be obtained,
\begin{eqnarray}
    H(t)=&&\left[i\sum_{i=1}^L\braket{v_i(t)|\dot{v}_{L+1}(t)}\ket{v_i(t)}\bra{v_{L+1}(t)}+{\rm h.c.}\right]\nonumber\\&&+\left[i\braket{v_{L+1}|\dot{v}_{L+1}(t)}-\dot{\gamma}(t)\right]
    \ket{v_{L+1}}\bra{v_{L+1}}\nonumber\\&&\nonumber+i\sum_{i\neq j}^L\braket{v_j(t)|\dot{v}_i(t)}\ket{v_j(t)}\bra{v_i(t)}\\&&-i\eta\sum_{j=1}^L\braket{v_j|\dot{v_j}}\ket{v_j}\bra{v_j}\label{eq:9}.
\end{eqnarray}
The derivation of the Eq.~(\ref{eq:9}) is provided in the appendix A. Eq.~(\ref{eq:9}) can be viewed as as a reverse design method. When the evolution trajectory of the computational subspace is given, we can determine the exact Hamiltonian that satisfies the UNHQC+ conditions Eq.~(\ref{eq:5}) and Eq.~(\ref{eq:6}).
\section{Application}
In Section \uppercase\expandafter{\romannumeral 2}, the UNHQC+ is proposed to reduce the time of quantum evolution, and a general  reverse engineering approach is provided for realizing the UNHQC+.
A key advantage of this method is its ability to achieve quantum gates along arbitrary paths. As an example, we will construct arbitrary sigle-qubit gate and the nontrivial two-qubit gate.

\subsection{One-qubit gate}
We consider a three-dimensional Hilbert space spanned by the states $\{\ket{0},\ket{1},\ket{e}\}$ to realize quantum gates. The general form of our auxiliary  states can be written as
\begin{eqnarray}
    \ket{v_1(t)}=&&\cos{\frac{\theta}{2}}\ket{0}+\sin{\frac{\theta}{2}}e^{i\varphi}\ket{1},\nonumber\\
    \ket{v_2(t)}=&&\cos{\frac{\chi(t)}{2}}\left[\sin{\frac{\theta}{2}}e^{-i\varphi}\ket{0}-\cos{\frac{\theta}{2}}\ket{1}\right]\nonumber\\&&+\sin{\frac{\chi(t)}{2}}e^{i\xi(t)}\ket{e},\nonumber\\
    \ket{v_3(t)}=&&\sin{\frac{\chi(t)}{2}}e^{-i\xi(t)}\left[\sin{\frac{\theta}{2}}e^{-i\varphi}\ket{0}-\cos{\frac{\theta}{2}}\ket{1}\right]\nonumber\\&&-\cos{\frac{\chi(t)}{2}}\ket{e},\label{eq:10}
\end{eqnarray}
where the $\ket{v_1(t)}$ is the dark state, and $\theta$, $\varphi$ are time-independent. The computational subspace is spanned by $\{v_1(t),v_2(t)\}$
and the $\chi(t)$ and $\xi(t)$ are time-dependent with $\chi(0)=\chi(T)=0$ to ensure $\ket{v_2(0)}=\ket{v_2(T)}$.
Substituting the states into Eq.~(\ref{eq:9}), we obtain
\begin{eqnarray}
H(t)=&&\frac{1}{2}\bigg\{e^{i\left[\varphi+\xi(t)+\Gamma(t)\right]}\Omega(t)\sin{\frac{\theta}{2}}\ket{e}\bra{0}\nonumber\\
&&+e^{i\left[\xi(t)+\pi+\Gamma(t)\right]}\Omega(t)\cos{\frac{\theta}{2}}\ket{e}\bra{1}+h.c.\bigg\}\nonumber\\
&&\Delta(t)\ket{e}\bra{e},\label{eq:11}
\end{eqnarray}
where we have set $\dot{\gamma}(t)=[3+\eta+(1+\eta)\cos{\chi(t)}]\xi(t)$ to decouple states $\ket{0}$ and $\ket{1}$. The parameters in Eq.~(\ref{eq:11}) read
\begin{eqnarray}
    \Omega(t)=&&\sqrt{\dot{\chi}^2(t)+\left[(1+\eta)\dot{\xi}(t)\sin{\chi(t)}\right]^2}\nonumber\\
    \Delta(t)=&&-\left[1+(1+\eta)\cos{\chi(t)}\right]\dot{\xi}(t)\nonumber\\
    \Gamma(t)=&&\arctan\left\{\dot{\chi}(t)/\left[(1+\eta)\dot{\xi}(t)\sin{\chi}(t)\right]\right\}.
\end{eqnarray}
The Hamiltonian Eq.~(\ref{eq:11}) describes a three-level system driven by a pair of Rabi pulses,
i.e., $\Omega(t)\sin{(\theta/2)}$ and $\Omega(t)\cos{(\theta/2)}$, in the two-photon resonant way with the detuning $\Delta(t)$. We can verify that the Hamiltonian Eq.~(\ref{eq:11}) and the auxiliary states Eq.~(\ref{eq:10}) satisfy the constraints given by Eq.(5) and (6). The elements of
the geometric matrix $A(t)$ can then be obtained
\begin{equation}
    A(t)=\begin{pmatrix}
            0&0\\
            0&-\sin^2{\frac{\chi}{2}}\dot{\xi}
         
        \end{pmatrix}
    \end{equation}
where we omit $(t)$ dependent for simplicity. The dynamical matrix $K(t)$ is written as
\begin{equation}
    K(t)=\begin{pmatrix}
            0&0\\
            0&\eta\sin^2{\frac{\chi}{2}}\dot{\xi}
        \end{pmatrix}
    \end{equation}
We notice directly  that $K_{22}=-\eta A_{22}$ and  $K_{ij}=A_{ij}$ =0 for $i\neq j$. The reason is that  the auxiliary state $\ket{v_1}$ in Eq.~(\ref{eq:10}) is independent of time. Hence, we always have $A_{1i}=\braket{v_1|i\partial_t|v_i}=0$. Of course, our reverse engineering does not require the auxiliary state to be time-independent. 
For example, The variables $\{\varphi(t),\chi(t)\}$ can be set to be time-dependent and  the variables $\{\theta,\xi(t)\}$ are time-independent , making all three auxiliary states time-dependent. The result is shown in the appendix B. Finally, the evolution operator is given by
\begin{equation}
    U(T) = e^{i\phi(T)\textbf{n}\bm{\sigma}/2},
\end{equation}
where $\phi(T)=\int_{0}^{T}A_{22}+k_{22}dt=\int_{0}^{T} (1+\eta) \sin^2{\frac{\chi(t)}{2}}\dot{\xi}(t)dt$ , $\textbf{n}=(\sin\theta\cos\varphi,\sin\theta\sin\varphi,\cos\theta)$,
and $\bm{\sigma}$ is the Pauli vector. We note that $\phi$ is equal to $(1+\eta)/2$ of the solid angle corresponding to the closed evolution path on the Bloch sphere.

\begin{figure}[]
    \centering
    \includegraphics[width=1\linewidth]{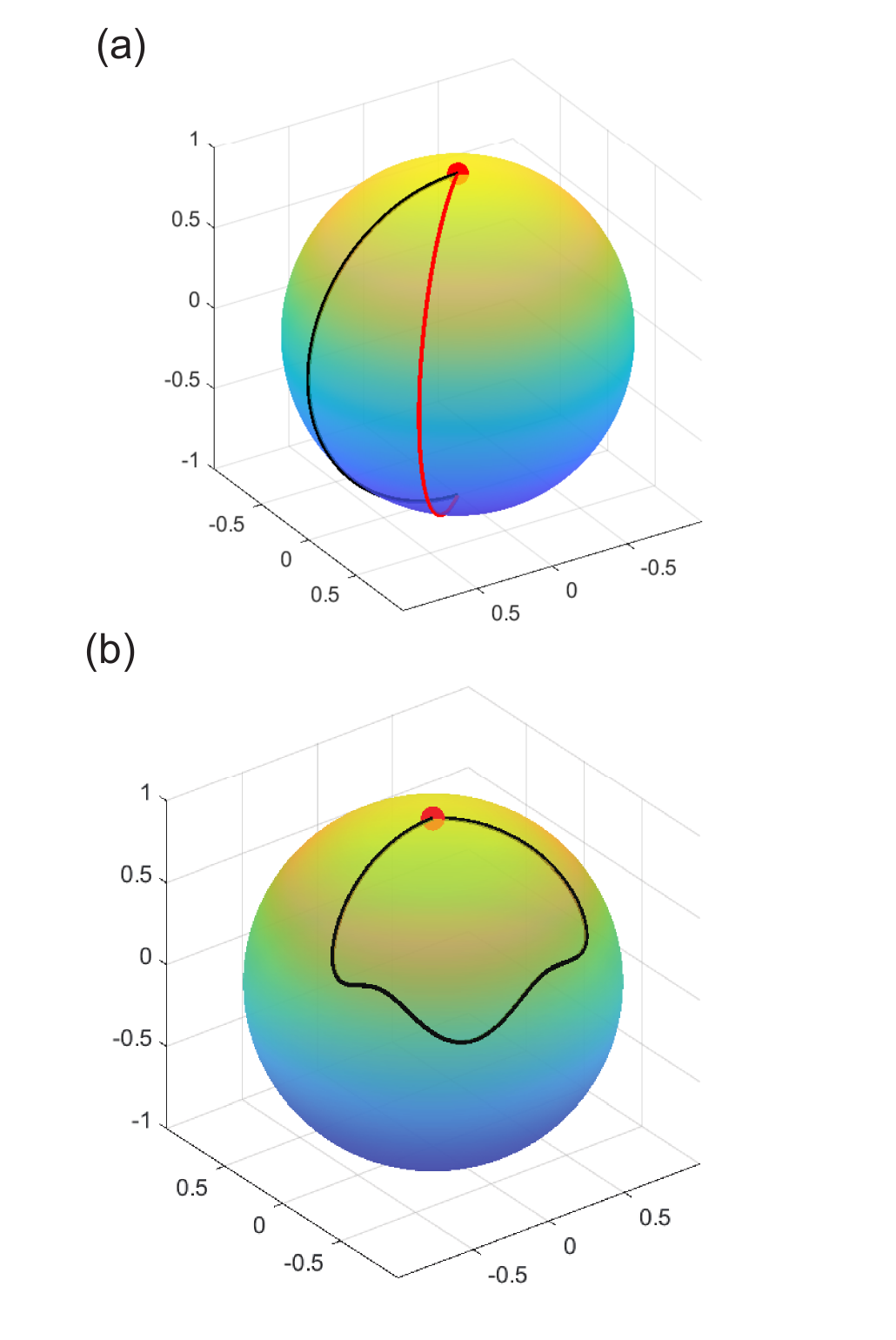}
    \caption{\label{fig:fig1}Evolution paths of the first and second strategy. (a) shows the conventional OSSP loop, where the quantum state evolves from the North Pole to the South Pole along the red geodesic, $\xi(t)=0$, and then returns from the South Pole to the North Pole along the black geodesic, $\xi(t)=\pi/4$. (b) shows the evolution path of the UNHQC+ strategy. }
\end{figure}

The parameters $\xi(t)$ and $\chi(t)$ are two undetermined functions, which depend on the evolution path on the Bloch sphere. Once the path is determined, the path parameters $\xi(t)$ and $\chi(t)$ are uniquely determined, 
and we can obtain the corresponding Hamiltonian. To illustrate this, we build the  T-gate, i.e., $e^{i\pi/8\cdot\sigma_z}$, with $(\theta,\varphi,\phi)=(0,0,\pi/4)$.\par
In the first strategy, as shown in Fig.\ref{fig:fig1}(a), the conventional OSSP method can be realized. We take $(\chi(t), \xi(t))$ starting from the north pole $(0,0)$ along the geodesic line with $\xi(t) = 0$ to the south pole $(\pi,0)$,
and then return to the north pole from the south along another geodesic line $\xi(t) = \pi/4$. The total phase reads $\phi(T)=\triangle \xi=\pi/4$.
On the first geodesic line, the Rabi frequency reads $\Omega(t)=|\dot{\chi}(t)|$, the detuning $\Delta(t)$ is always zero, and the phase is $\Gamma(t)=\pi/2$. For the second geodesic line, the form of the Rabi frequency and the detuning remain unchanged, but the phase is $\Gamma(t)=-\pi/2$.
The path length of this process is $l_0 = 2\pi$. We find that regardless of the rotation angle of the quantum gate, the path length is always $2\pi$, which is a weakness that is difficult to overcome in OSSP theory. Additionally, in OSSP, the quantum state requires an angular mutation at the South Pole, which also contributes to the decrease in the fidelity of the quantum gate.\par

In the UNHQC+ strategy, we use the parameter fitting method to find a faster evolution solution and a smoother path to improve the robustness of quantum gates. In first step, taking the Fourier form of the angle $\mu\in (\chi,\xi)$\cite{liang2024nonadiabatic},
\begin{equation}
    \mu(t) = \sum_{n=1}a_n^\mu\bigg[\sin(\frac{b_n^\mu\pi t}{T})\bigg]^{c_n^\mu}. \label{eq:16}
\end{equation}
where $a_n^v,b_n^v,c_n^v$ are the free parameters. We can substitute $\mu(t)$ into $\phi(t)$, i.e., $\phi(T)=\int_{0}^{T}(1+\eta)\sin^2{\frac{\chi(t)}{2}}\dot{\xi}(t)dt$. Then, the objective function is determined according to the rotation angle $\phi(T)$ of the quantum gate that you want to build. For example, we need $\phi(T)=\pi/4$ in the T-gate. The free variables can be obtained by fitting $\phi(T)=\pi/4$.\par
\begin{figure}[]
    \centering
    \includegraphics[width=1\linewidth]{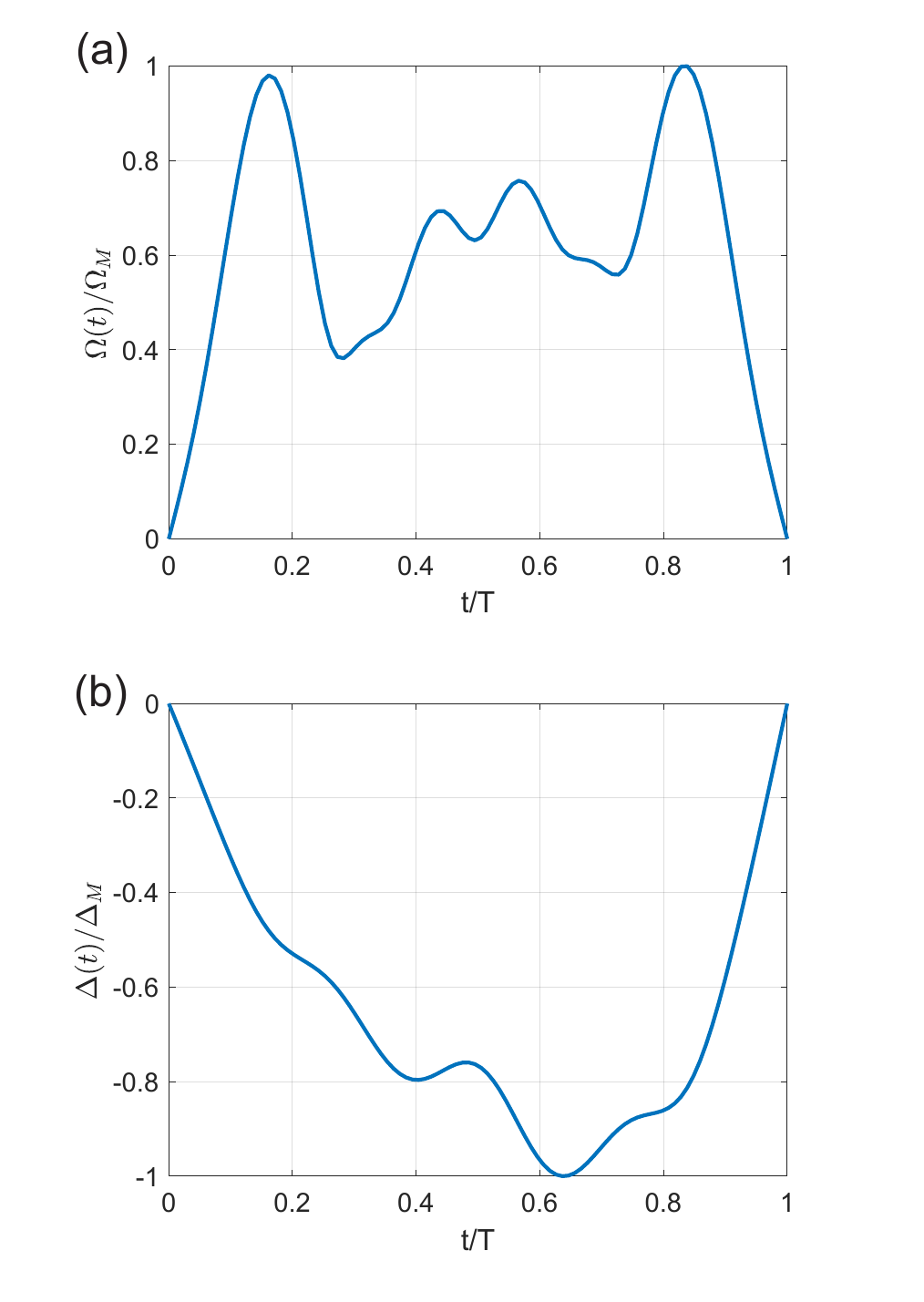}
    \caption{\label{fig:figpulse} The Rabi frequency and detuning of our UNHQC+ scheme. Here, (a) shows the pulse shape of Rabi frequency $\Omega(t)$ and (b) is the shape of detunning $\Delta(t)$.}
\end{figure}
We present a detailed procedure for the construction of a T-gate. First, the free parameters are set as $b_n^\chi=n$, $c_n^\chi=2$, and $b_n^\xi=1/2$, $c_n^\xi=n+1$, to ensure that the quantum states return to their initial position after periodic evolution.
It should be noted that the value ranges of $\eta$ and $n$ can be set arbitrarily in our reverse engineering. We take $n\in{1,2,3,4}$ and $\eta=0.2$ as an example. The solutions reads 
\begin{eqnarray}
    \{a_1^\chi,a_2^\chi,a_3^\chi,a_4^\chi\} =&&\{-1.22, -0.39, -0.15,0.14\}\nonumber\\
    \{a_1^\xi,a_2^\xi,a_3^\xi,a_4^\xi\} =&&\{1.82, 0.65, -0.96,1.05\}.\nonumber\\ \label{eq:17}  
\end{eqnarray}
Figure \ref{fig:fig1}(b) presents the corresponding evolution path of the quantum state under the control of this parameter set. The length of the UNHQC+ evolution curve in fig.\ref{fig:fig1}(b)  is $0.77\pi$. Substituting Eq.~(\ref{eq:16}) and Eq.~(\ref{eq:17}) into reverse engineering  Eq.~(\ref{eq:11}), the system parameters $\Omega(t)$ and $\Delta(t)$ are obtained. The profile of pulse is shown in Fig.\ref{fig:figpulse} with the amplitude of Rabi pulse $\Omega_M=2\pi\times 20$ MHz. The evolution time of UNHQC+ is $57.6$ns. In the following, we set the Rabi pulse of the OSSP scheme as $\Omega(t)=\Omega_M\sin(\pi t/T)^2$, with the same amplitude $\Omega_M$. The calculation shows that the evolution time of the OSSP is about $100$ns. As shown, this method leads to a considerable reduction in evolution time.   \par
\begin{figure}[]
    \centering
    \includegraphics[width=1\linewidth]{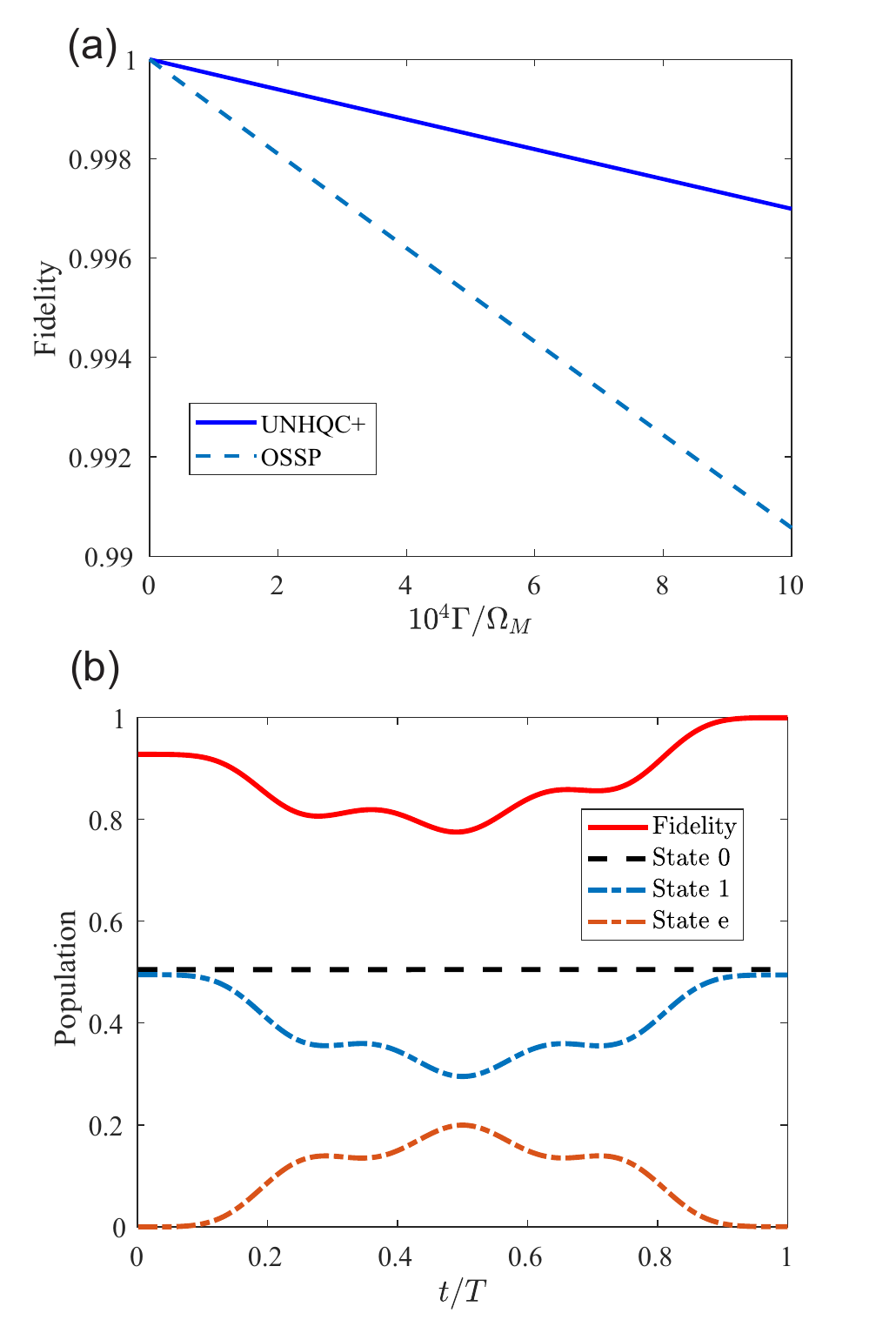}
    \caption{\label{fig:fig2}(a)T-gate fidelity for OSSP (dashed line) and UNHQC+ (solid line) as function of decoherence rate. (b)Dynamics of the gate fidelity and states population for UHNQC+ T-gate, with the fidelity reaching $99.94\%$.}
\end{figure}

The Lindblad master equation\cite{lindblad1976generators} is used to evaluate the UNHQC+ and OSSP T-gate robustness
\begin{eqnarray}
    \dot{\rho}_1=-i[H(t),\rho_1]+\frac{1}{2}\sum_{j=1,2}\Gamma_j L(\sigma_j)
\end{eqnarray}
where $\rho$ is the quantum density matrix and $L(A)=2A\rho A -A^\dagger A-\rho A^\dagger A $ is the Lindbladian operator with $\sigma_1=\ket{0}\bra{e}+\ket{1}\bra{e}$ and $\sigma_2=\ket{e}\bra{e}-\ket{1}\bra{1}-\ket{0}\bra{0}$ . $\Gamma_{1,2}$ represent the decay and dephasing rate. 
We estimate average fidelity as $F_G=\int_{0}^{2\pi} \bra{\nu_i(\tau)}\rho_1\ket{\nu_i(\tau)} \,d\theta /2\pi$, with the integration performed numerically for 100 initial states where $\theta$ is evenly 
distributed over $[0,2\pi]$, the initial state reads $\ket{\nu_i(0)}=\cos{\theta_i}\ket{0}+\sin{\theta_i}\ket{1}$, and the perfect evolution state shows $\ket{\nu_i(\tau)}=U(\tau)\ket{\nu_i(0)}$. Fig.\ref{fig:fig2}(a) shows the reduction in gate fidelity due to decoherence for the OSSP and UNHQC+ strategies, with $\Gamma_1=\Gamma_2$. We find that the fidelity of the UNHQC+ method is consistently higher than that of the OSSP across different decoherence and dephasing rates. we also set $\Gamma_1=\Gamma_2=2\pi\times4$ kHz to evaluate the state populations and the fidelity of UNHQC+. The result is shown in Fig.\ref{fig:fig2}(b), where the gate fidelity is about $99.94\%$.\par

\begin{figure}[]
    \centering
    \includegraphics[width=1\linewidth]{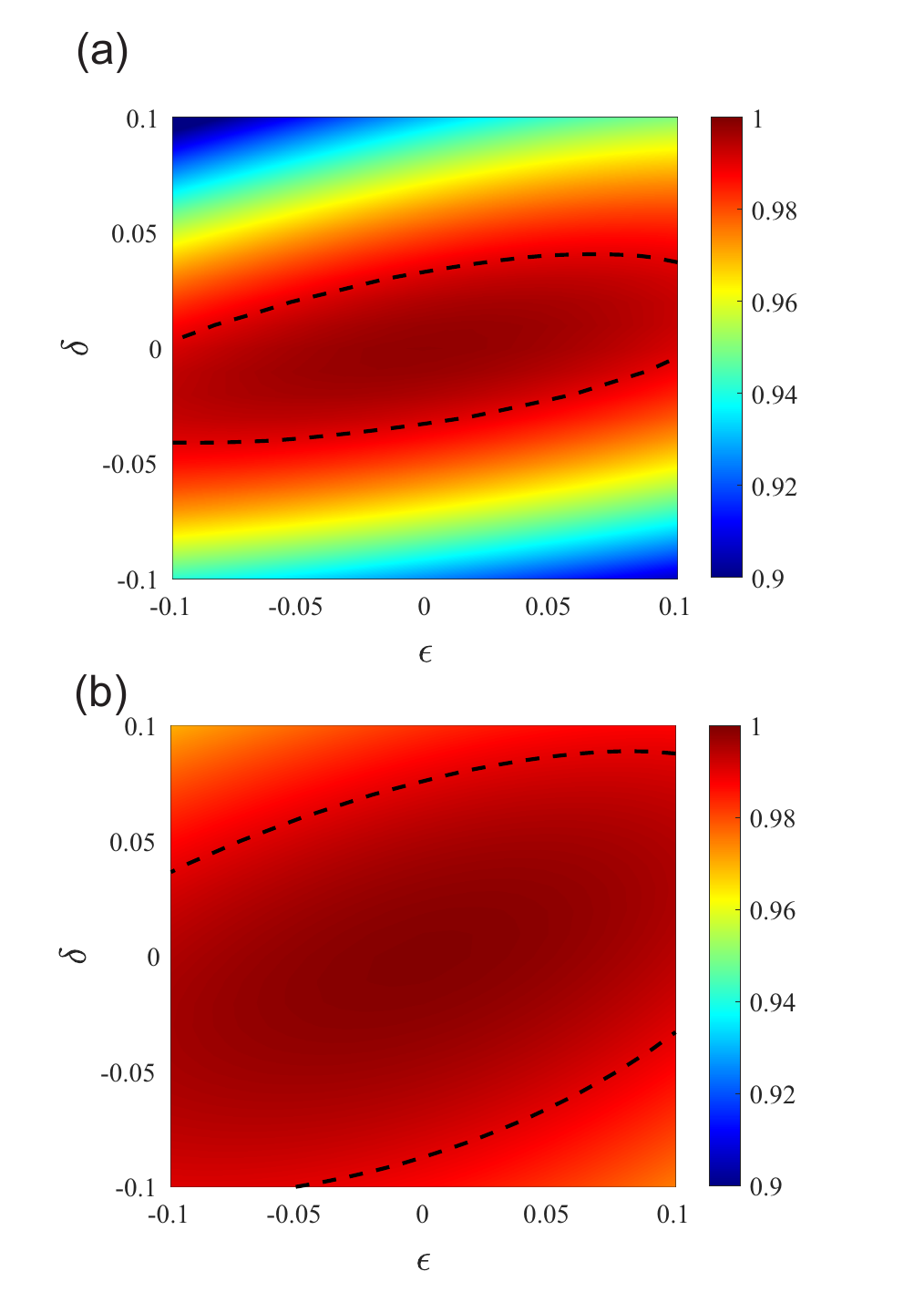}
    \caption{\label{fig:fig3}Gate fidelities  of  (a)OSSP and (b)UNHQC+ T-gates as functions of control errors $\delta$ and $\epsilon$.  Here the black dashed line represents the fidelity $99\%$.}
\end{figure}

The fidelity of the UNHQC+ method is also more robust to control errors than the OSSP method. Here, we consider the detuning error and Rabi error induced by imperfect control based on the ideal Hamiltonian, in the form of  $\Delta\to\Delta+\delta\Omega_M$ and $\Omega\to(1+\epsilon)\Omega$. Fig.\ref{fig:fig3} (a) shows the fidelity of the OSSP T-gate as a function of control error and Fig.\ref{fig:fig3}(b) is the fidelity of the UNHQC+ T-gate, where the black dashed line represents the fidelity $99\%$.
\subsection{Two-qubit gate}
To realize an UNHQC+ nontrivial two-qubit gate, we select the states as
\begin{eqnarray}
        \ket{v_1(t)}=&&\ket{00},\ket{\psi_2(t)}=\ket{01},\nonumber\\
        \ket{v_3(t)}=&&\cos{\frac{\theta}{2}}\ket{10}+\sin{\frac{\theta}{2}}e^{i\varphi}\ket{11},\nonumber\\
        \ket{v_4(t)}=&&\cos{\frac{\chi(t)}{2}}\left[\sin{\frac{\theta}{2}}e^{-i\varphi}\ket{10}-\cos{\frac{\theta}{2}}\ket{11}\right]\nonumber\\&&+\sin{\frac{\chi(t)}{2}}e^{i\xi(t)}\ket{ee},\nonumber\\
        \ket{v_5(t)}=&&\sin{\frac{\chi(t)}{2}}e^{-i\xi(t)}\left[\sin{\frac{\theta}{2}}e^{-i\varphi}\ket{10}-\cos{\frac{\theta}{2}}\ket{11}\right]\nonumber\\&&-\cos{\frac{\chi(t)}{2}}\ket{ee},\nonumber\\\label{eq:19}
\end{eqnarray}
where $\chi(0)=\chi(T)=0$. The computational subspace is spanned by $\{v_i(t)\}_{i=1}^4$. We directly substitute  Eq.~(\ref{eq:19}) into  Eq.~(\ref{eq:9}) and obtain the Hamiltonian
\begin{eqnarray}
H(t)=&&\frac{1}{2}\bigg\{e^{i\left[\varphi+\xi(t)+\Gamma(t)\right]}\Omega(t)\sin{\frac{\theta}{2}}\ket{ee}\bra{10}\nonumber\\
&&+e^{i\left[\xi(t)+\pi+\Gamma(t)\right]}\Omega(t)\cos{\frac{\theta}{2}}\ket{ee}\bra{11}+h.c.\bigg\},\nonumber\\
&&\Delta(t)\ket{ee}\bra{ee}
\end{eqnarray}
The control parameters read
\begin{eqnarray}
    \Omega(t)=&&\sqrt{\dot{\chi}^2(t)+\left[(1+\eta)\dot{\xi}(t)\sin{\chi(t)}\right]^2}\nonumber\\
    \Delta(t)=&&-\left[1+(1+\eta)\cos{\chi(t)}\right]\dot{\xi}(t)\nonumber\\
    \Gamma(t)=&&\arctan\left\{\dot{\chi}(t)/\left[(1+\eta)\dot{\xi}(t)\sin{\chi}(t)\right]\right\}.
\end{eqnarray}
In this case, after a cyclic evolution, the unitary operator
acting on the computational space shows $U(T)=U_{s1}(T)\oplus U_{s2}(T) $, where $U_{s1}(T)=\ket{0}\bra{0}\otimes I$.
and 
\begin{eqnarray}
    U_{s2}(T)=&& e^{-i\phi(T)}\ket{\psi_4(0)}\bra{\psi_4(0)} + \ket{\psi_3(0)}\bra{\psi_3(0)}\nonumber\\
            =&&e^{-i\phi(T)/2}\ket{1}\bra{1}\otimes e^{i\phi(T)\textbf{n}\bm{\sigma}/2},
\end{eqnarray}
where $\phi(T)=\int_{0}^{T} (1+\eta) \sin^2{\frac{\chi(t)}{2}}\dot{\xi}(t)dt$. We find the general form of the Hamiltonian here, which is similar to that in Eq.~(\ref{eq:11}). Therefore, the discussion on the selection of robustness paths for two-qubit gates is the same as for single-qubit gates.
\section{conclusion}
Recently, an optimized holonomic quantum computation (NHQC+) approach has been proposed to improve the flexibility of NHQC. We combine this optimized approach with an unconventional geometric scheme to enhance the fidelity and robustness of quantum gates. Of course, this comes with challenges, such as the need for more complex control parameter design and evolution loops. Here, we present a general reverse-engineering method to address this problem. Our reverse engineering is flexible because it does not impose restrictions on the choice of quantum states or path parameters. Leveraging this advantage, we can select faster and more robust evolution paths. Guided by this principle, we have designed a faster and more robust T-gate evolution loop as a demonstrative example The results show that our method is more robust against control errors and decoherence than the conventional OSSP method. The examples presented in this paper demonstrate the broad applicability of our method, which can be extended to many other scenarios beyond those discussed here.

\section{ACKNOWLEDGMENTS}
The work is supported by National Natural Science Foundation of China Grant No.62588201, The Key Research and Development Program of Guangdong Province Grant No. 2019B090917007, the Science and Technology Planning Project of Guangdong Province Grant No.2019B090909011 and Guangdong Provincial Key Laboratory Grant No.2019B121203002. Qi Li was supported by National Natural Science Foundation of China Grant
No.12504174.

\appendix

\section{derivation of reverse engineering}
We provide the derivation of Eq.~(\ref{eq:9}). The first step expands the time derivative operator,
\begin{eqnarray}
    H =&&i\sum_{i,j,m}^L\left[\dot{C}_{ji}C_{im}^*\ket{v_{j}}\bra{v_m}+C_{ji}C_{im}^*\ket{\dot{v}_j}\bra{v_m}\right]\nonumber\\
    &&+i\left[i\dot{\gamma}\ket{v_{L+1}}+\ket{\dot{v}_{L+1}}\right]\bra{v_{L+1}}.
\end{eqnarray}
Substituting Eq.~(\ref{eq:7}) into above $H$, we obtain
\begin{eqnarray}
    H =&&i\sum_{i,j,m}^L\left[-(1+\eta)\braket{v_j|\dot{v}_j}C_{im}^*\ket{v_{j}}\bra{v_m}+C_{ji}C_{im}^*\ket{\dot{v}_j}\bra{v_m}\right]\nonumber\\
    &&+i\left[i\dot{\gamma}\ket{v_{L+1}}+\ket{\dot{v}_{L+1}}\right]\bra{v_{L+1}}.
\end{eqnarray}
Because matrix $C$ is the unitary matrix, i.e. $\sum_i^L C_{ji}C_{im}^*=\delta_{jm}$, the Hamiltonian can be written as
\begin{eqnarray}
    H =&&i\sum_{j}^L\left[-(1+\eta)\braket{v_j|\dot{v}_j}\ket{v_{j}}\bra{v_j}+\ket{\dot{v}_j}\bra{v_j}\right]\nonumber\\
    &&+i\left[i\dot{\gamma}\ket{v_{L+1}}+\ket{\dot{v}_{L+1}}\right]\bra{v_{L+1}}.
\end{eqnarray}
Substituting $\sum_{i=1}^{L+1}\ket{v_i}\bra{v_i}=1$ into $H$ and using $\braket{v_i|v_j}=\delta_{ij}$,
\begin{eqnarray}
    H =&&i\sum_{j=1}^L\left[-(1+\eta)\braket{v_j|\dot{v}_j}\ket{v_{j}}\bra{v_j}\right]
    +i\sum_{j=1}^L\sum_{i=1}^{L+1}\braket{v_i|\dot{v}_j}\ket{v_i}\bra{v_j}\nonumber\\
    &&-\dot{\gamma}\ket{v_{L+1}}\bra{v_{L+1}}+i\sum_{i=1}^{L+1}\braket{v_i|\dot{v}_{L+1}}\ket{v_{i}}\bra{v_{L+1}}.
\end{eqnarray}
The summation in the second term of $H$, $\sum_{i=1}^{L+1}$, can be decomposed as $\sum_{i=j} + \sum_{i \neq j}$. This decomposition reveals that the second term partially cancels with the first term, leading to
\begin{eqnarray}
    H =&&\sum_{j=1}^L\left[-i\eta\braket{v_j|\dot{v}_j}\ket{v_{j}}\bra{v_j}\right]
    +i\sum_{i\neq j}^L\braket{v_i|\dot{v}_j}\ket{v_i}\bra{v_j}\nonumber\\
    &&+i\sum_{j=1}^L\braket{v_{L+1}|\dot{v}_j}\ket{v_{L+1}}\bra{v_j}+i\sum_{i=1}^{L}\braket{v_i|\dot{v}_{L+1}}\ket{v_{i}}\bra{v_{L+1}}
   \nonumber\\
    && +(i\braket{v_{L+1}|\dot{v}_{L+1}}-\dot{\gamma})\ket{v_{L+1}}\bra{v_{L+1}}.
\end{eqnarray}
Finally, the third and forth terms in above $H$ are Hermitian conjugates. Therefore,
\begin{eqnarray}
    H(t)=&&\left[i\sum_{i=1}^L\braket{v_i(t)|\dot{v}_{L+1}(t)}\ket{v_i(t)}\bra{v_{L+1}(t)}+{\rm h.c}\right]\nonumber\\&&+\left[i\braket{v_{L+1}|\dot{v}_{L+1}(t)}-\dot{\gamma}(t)\right]
    \ket{v_{L+1}}\bra{v_{L+1}}\nonumber\\&&\nonumber+i\sum_{i\neq j}^L\braket{v_j(t)|\dot{v}_i(t)}\ket{v_j(t)}\bra{v_i(t)}\\&&-i\eta\sum_{j=1}^L\braket{v_j|\dot{v_j}}\ket{v_j}\bra{v_j}.
\end{eqnarray}

\section{three time-independent states}
Here, the variables $\{\varphi(t), \chi(t)\}$ are time dependent and $\{\theta, \xi\}$ are not. The auxiliary states can be rewritten as
\begin{eqnarray}
    \ket{v_1(t)}=&&\cos{\frac{\theta}{2}}\ket{0}+\sin{\frac{\theta}{2}}e^{i\varphi(t)}\ket{1},\nonumber\\
    \ket{v_2(t)}=&&\cos{\frac{\chi(t)}{2}}\left[\sin{\frac{\theta}{2}}e^{-i\varphi(t)}\ket{0}-\cos{\frac{\theta}{2}}\ket{1}\right]\nonumber\\&&+\sin{\frac{\chi(t)}{2}}e^{i\xi}\ket{e},\nonumber\\
    \ket{v_3(t)}=&&\sin{\frac{\chi(t)}{2}}e^{-i\xi}\left[\sin{\frac{\theta}{2}}e^{-i\varphi(t)}\ket{0}-\cos{\frac{\theta}{2}}\ket{1}\right]\nonumber\\&&-\cos{\frac{\chi(t)}{2}}\ket{e},
\end{eqnarray}
In order to satisfy the cyclic evolution condition, we need to set $\chi(T)=\chi(0)=0$ and $\varphi(0)=\varphi(T)$. Substituting them into Eq.~(\ref{eq:9}), the Hamitonian is obtained.
\begin{eqnarray}
H(t)=&&\left\{\Omega_1\ket{0}\bra{e}+\Omega_2\ket{1}\bra{e}+{\rm h.c.}\right\}
+\Delta_1\ket{0}\bra{0}
\nonumber\\&&+\Delta_2\ket{1}\bra{1},
\end{eqnarray}
It means that $\ket{0}$ and $\ket{e}$ are driven by a laser with Rabi $\Omega_1$ and the detuning $\Delta_1$, meanwhile $\ket{1}$ and $\ket{e}$ are driven by a laser with Rabi $\Omega_2$ and the detuning $\Delta_2$.
Because the parameters $\Omega_{1,2}$ and $\Delta_{1,2}$ here are complex, for simplicity we only give the form of geometric matrix $A$ and dynamic matrix $K$,
\begin{equation}
    A(t)=\begin{pmatrix}
            -\dot{\varphi}(t)\sin^2{\frac{\theta}{2}}&e^{-i\varphi(t)}\frac{\dot{\varphi}(t)}{2}\sin\theta\cos\frac{\chi(t)}{2}\\
            e^{i\varphi(t)}\frac{\dot{\varphi}(t)}{2}\sin\theta\cos\frac{\chi(t)}{2}&\dot{\varphi}(t)\sin^2{\frac{\theta}{2}}\cos^2{\frac{\chi(t)}{2}}
    \end{pmatrix}
\end{equation}

\begin{equation}
    K(t)=\begin{pmatrix}
            \eta\dot{\varphi}(t)\sin^2{\frac{\theta}{2}}&e^{-i\varphi(t)}\frac{\dot{\varphi}(t)}{2}\sin\theta\cos\frac{\chi(t)}{2}\\
            e^{i\varphi(t)}\frac{\dot{\varphi}(t)}{2}\sin\theta\cos\frac{\chi(t)}{2}&-\eta\dot{\varphi}(t)\sin^2{\frac{\theta}{2}}\cos^2{\frac{\chi(t)}{2}}
    \end{pmatrix}
\end{equation}
It is obvious that the two matrices satisfy the conditions Eq.~(\ref{eq:5}) and Eq.~(\ref{eq:6}).
\nocite{*}

\bibliography{ref}
\end{document}